\documentstyle{article}
\begin{document}
\title {
\bf \huge
Integrable boundary impurities in the $t-J$ model
with different gradings
}

\author{
{\bf Heng Fan$^{a,b}$, Miki Wadati$^a$}\\
$^a$\normalsize Department of Physics, Graduate School of Science,\\
\normalsize University of Tokyo, Hongo 7-3-1,\\
\normalsize Bunkyo-ku, Tokyo 113-0033, Japan.\\
\normalsize $^b$Institute of Modern Physics, P.O.Box 105,\\
\normalsize Northwest University, Xi'an 710069, P.R.China.
}

\maketitle

\begin{abstract}
We investigate the generalized supersymmetric $t-J$ model with
boundary impurities in different gradings.
All three different gradings: fermion, fermion, boson (FFB),
boson, fermion, fermion (BFF) and fermion, boson, fermion (FBF),
are studied for the generalized supersymmetric $t-J$ model.
Boundary  K-matrix operators are found for the different gradings.
By using the graded algebraic Bethe ansatz method,
we obtain the eigenvalues and the corresponding Bethe ansatz equations
for the transfer matrix.
\end{abstract}


\newpage
\baselineskip 0.5truecm

\section{Introduction}
There has been extensive interests in the investigation of
the impurity problems. The Anderson and Kondo lattice
models describe the physics of conduction electrons
in extended orbitals interacting with strongly correlated electrons
in localized orbitals (impurities), see Ref.\cite{TSU} for a review.
For superconductivity, the nonmagnetic impurities
are generally belived to have little effect on
the superfliud density and the transition temperature
in conventional superconductivity\cite{A}.
For the high-temperature cuprate superconductors,
both magnetic and nonmagnetic impurities,
for example, nonmagnetic impurity Zn and magnetic
impurity Ni in cuprate LSCO, play
an important role for the interpretation of experiments\cite{NP}.
 
The study of the effects due to the presence of impurities in
1-dimensional quantum chains in the framework of integrable
models has a long successful history\cite{AJ,AFL,TW}.
In the framework of the Quantum Inverse
Scattering Method (QISM),
Andrei and Johannesson\cite{AJ} studied an arbitrary
spin S embeded in a spin-1/2 Heisenberg chain by the
algebraic Bethe ansatz method\cite{KIB}.
This method was later generalized to other cases
including the integrable supersymmetric $t-J$ model
\cite{LS,BEF}.

The $t-J$ model proposed by Zhang and Rice\cite{ZR}
is one of the most widely accepted models to
describe the high-temperature Cu-oxide superconductors.
The Hamiltonian of the $t-J$ model
includes the nearest neighbour hopping term ($t$) and
the antiferromagnetic exchange term ($J$). The
Hamiltonian of 1-dimensional $t-J$ model is written as
\begin{eqnarray}
H=\sum _{j=1}^N\left\{ -t{\cal {P}}
\sum _{\sigma =\pm 1}(c_{j, \sigma}^{\dagger }
c_{j+1, \sigma }+c_{j+1,\sigma }^{\dagger }c_{j,\sigma} ){\cal {P}}
+J({\bf S}_j{\bf S}_{j+1}                      
-{1\over 4}n_nn_{j+1})\right\},
\end{eqnarray}
where the Gutzwiller's projection operator
$\cal{P}$ ensures the exclusion of double occupancy in
one lattice site.
It is known that this model is supersymmetric and integrable
for $J=\pm 2t$ \cite{L,S}. The supersymmetric $t-J$ model was also
studied in Refs.\cite{S2,EK,FK}.
Using the graded algebraic Bethe ansatz method, the integrable
supersymmetric $t-J$ model was studied for all three different
gradings\cite{EK} for periodic boundary conditions.
The generalized supersymmetric $t-J$ model with open
boundary conditions for three different gradings
is studied in Ref.\cite{FWW}.

Traditionally, the integrable models are
studied for periodic boundary conditions where the Yang-Baxter
equation play a key role\cite{YB}.
In the last decade, the study of integrable models with open
(reflecting) boundary conditions have been attracting a great
deal of interests\cite{S3}. Besides the Yang-Baxter
equation, the reflection equations are also important
for the study of the open boundary conditions.
Applying the method used in \cite{AJ} and
the reflection equation, the Heisenberg
spin chain with boundary impurities is studied\cite{W}.
And the same method is applied to
the integrable supersymmetric $t-J$ model\cite{WDHP,ZGLG},
see also \cite{BF} for related works.
The boundary impurities for the generalized (q-deformed)
$t-J$ model are also studied by
the graded algebraic Bethe ansatz method\cite{FWY,GGLZ}.
In the previous works \cite{WDHP,ZGLG,FWY,GGLZ}, the
boundary impurities refer to the boundary impurity spins
coupled with the original $t-J$ spin chain,
so this impurities should correspond to the magnetic impurities.
In this paper, we shall extend our previous results in
\cite{FWY}. Besides the FFB grading, we shall deal
with BFF and FBF gradings. It seems that this kind of
boundary impurities corrrespond to the nonmagnetic
impurities. Thus all three possible gradings
are studied for the generalized $t-J$ model with
boundary impurities. Using the projecting method,
integrable $t-J$ model with boundary impurities for different gradings are
also studied in Ref.\cite {BF}.

The paper is organized as follows: We introduce the model in Section 2.
In section 3, we present briefly the results of FFB grading.
In section 4, using the nested algebraic Bethe ansatz method,
we obtain the eigenvalues of the transfer matrix with
grading BFF. In section 5, results for FBF grading are presented.
Section 6 includes a brief summary and discussions.

\section{Description of the Model}
In this paper, we shall study the generalized (q-deformed)
supersymmetric $t-J$ model.
The Hamiltonian of the model takes the following form:
\begin{eqnarray}
H&=&\sum _{j=1}^N\sum _{\sigma =\pm }
[\tilde {c}_{j,\sigma }^{\dagger}\tilde {c}_{j+1,\sigma }
+\tilde {c}_{j+1,\sigma }^{\dagger}
\tilde {c}_{j,\sigma }]
-2\sum _{j=1}^N[{1\over 2}(S_j^{\dagger }S_{j+1}+S_jS_{j+1}^{\dagger })
+cos(\eta )S_j^zS_{j+1}^z
-{cos(\eta )\over 4}n_jn_{j+1}]
\nonumber \\
&&+isin(\eta )\sum _{j=1}^N[S_j^zn_{j+1}-S_{j+1}^zn_j].
\end{eqnarray}
where $\tilde {c}_{j,\sigma }^{\dagger}
=(1-n_{j,-\sigma })c_{j,\sigma }^{\dagger},
\tilde {c}_{j,\sigma }=c_{j,\sigma }(1-n_{j,-\sigma})$.
When the anisotropic parameter $\eta =0$, this Hamiltonin
reduces to an equivalent form of the original Hamiltonian (1)
with $J=2t$.
The operators $c_{j,\sigma }$ and $c_{j,\sigma }^{\dagger }$ are
annihilation and creation operators of electron with spin $\sigma $
on a lattice site $j$, and we assume the total number of
lattice sites is $N$, $\sigma =\pm $ represent
spin down and up, respectively. These operators are canonical
Fermi operators satisfying anticommutation relations
$\{c_{j,\sigma }^{\dagger },c_{j,\tau }\} =\delta _{ij}
\delta _{\sigma \tau }$.
We denote by $n_{j,\sigma }=c_{j,\sigma }^{\dagger }c_{j,\sigma }$
the number operator for the electron on a site $j$ with
spin $\sigma $, and by
$n_j=\sum _{\sigma =\pm }n_{j,\sigma }$ the number operator for the
electron on a site $j$.
The Fock vacuum state $|0>$ is defined as $c_{j,\sigma }|0>=0$.
Due to the exclusion of double occupancy,
there are
altogether three possible electronic states at a given lattice site
$j$, two are fermionic and one is bosonic,
\begin{eqnarray}
|0>,~~~|\uparrow >_j=c_{j,1}^{\dagger }|0>,~~~
|\downarrow >_j=c_{j,-1}^{\dagger }|0>.
\end{eqnarray}
Here 
$S^z_j,S_j,S_j^{\dagger }$ are spin operators satisfying $su(2)$
algebra and can be expressed as
\begin{eqnarray}
S_j=c_{j,1}^{\dagger }c_{j,-1},
~~~~S_j^{\dagger }=c_{j,-1}^{\dagger }c_{j,1},
~~~~S_j^z={1\over 2}(n_{j,1}-n_{j,-1}).
\label{S}
\end{eqnarray}

The above Hamiltonian can be obtained from the logarithmic derivative
of the transfer matrix at
zero spectral parameter.
In the framework of QISM, the transfer matrix is constructed by
the trigonometric R-matrix of the Perk-Schultz
model \cite{PS}. The non-zero
entries of the R-matrix are given by
\begin{eqnarray}
&&{\tilde {R}}(\lambda )^{aa}_{aa}=sin(\eta +\epsilon _a\lambda ),
\nonumber \\
&&{\tilde {R}}(\lambda )^{ab}_{ab}=(-1)^{\epsilon _a\epsilon _b}
sin(\lambda ),
\nonumber \\
&&{\tilde {R}}(\lambda )^{ab}_{ba}=e^{isign(a-b)\lambda }sin(\eta ),
a\not= b,
\end{eqnarray}
where
$\epsilon _a$ is the Grassman parity,
$\epsilon _a=0$ for boson and $\epsilon _a=1$ for fermion, and
\begin{eqnarray}
sign(a-b)=\left\{
\begin{array}{ll}1,&{\rm if} ~a>b \\
-1, &{\rm if}~ a<b.
\end{array}\right.
\end{eqnarray}
This R-matrix satisfies the usual Yang-Baxter equation:
\begin{eqnarray}
{\tilde {R}}_{12}(\lambda -\mu )
{\tilde {R}}_{13}(\lambda )
{\tilde {R}}_{23}(\mu )
={\tilde {R}}_{23}(\mu )
{\tilde {R}}_{13}(\lambda )                                    
{\tilde {R}}_{12}(\lambda -\mu )
\end{eqnarray}
In this paper, we shall concentrate our discussion only on
the two fermion and one boson case, that means
for Grassmann parities $\epsilon _i,i=1,2,3$,
two of them equal to 1, and the last one equal to zero.
For example, we let
$\epsilon _1=\epsilon _2=1, \epsilon _3=0$ for FFB grading,
and $\epsilon _2=\epsilon _3=1, \epsilon _1=0$ for BFF grading.
We shall use
the graded formulae to study this model. For supersymmetric
$t-J$ model, the spin of the electrons and the charge `hole'
degrees of freedom play a very similar role forming a
graded superalgebra with two fermions and one boson. The holes
obey boson commutation relations, while the spinons are
fermions. The graded approach has an advantage
of making clear distinction between bosonic and fermionic degrees
of freedom \cite{OWA}.

Introducing a diagonal matrix
$\Pi _{ac}^{bd}=(-)^{\epsilon _a\epsilon _c}
\delta _{ab}\delta _{cd}$, we change the original R-matrix to the following
form,
\begin{eqnarray}
R(\lambda )=\Pi {\tilde {R}}(\lambda ).
\label{R}
\end{eqnarray}
From the non-zero elements of the R-matrix
$R_{ab}^{cd}$, we see that
$\epsilon _a+\epsilon _b+\epsilon _c+\epsilon _d=0$.
One can show that the R-matrix satisfies the graded Yang-Baxter equation,
\begin{eqnarray}
R(\lambda -\mu )_{a_1a_2}^{b_1b_2}
R(\lambda )_{b_1a_3}^{c_1b_3}
R(\mu )_{b_2b_3}^{c_2c_3}
(-)^{(\epsilon _{b_1}+\epsilon _{c_1})\epsilon _{b_2}}
=
R(\mu )_{a_2a_3}^{b_2b_3}R(\lambda )_{a_1b_3}^{b_1c_3}
R(\lambda -\mu )_{b_1b_2}^{c_1c_2}(-)^{(\epsilon _{a_1}
+\epsilon _{b_1})\epsilon _{b_2}}.
\label{GYBE}
\end{eqnarray}

In the framework of the QISM, we can construct
the $L$ operator from the R-matrix as
$L_{aq}(\lambda )=R_{aq}(\lambda )$, where the subscript $a$ represents
auxiliary space, and $q$ represents quantum space.
Thus we can rewrite the graded Yang-Baxter equation (\ref{GYBE})
as the following (graded) Yang-Baxter relation,
\begin{eqnarray}
R_{12}(\lambda -\mu )L_1(\lambda )L_2(\mu )
=L_2(\mu )L_1(\lambda )R_{12}(\lambda -\mu ).
\label{YBR}
\end{eqnarray}
Here the tensor product is in the sense of super-tensor product
defined as
\begin{eqnarray}
(F\otimes G)_{ac}^{bd}=F_a^bG_c^d(-)^{(\epsilon _a+\epsilon _b)\epsilon _c}.
\end{eqnarray}
Hereafter, all tensor products in this paper
are in the sense of super-tensor products.

It is standard that the row-to-row monodromy matrix
$T_N(\lambda )$ is defined as a matrix product over
the $N$ operators on all sites of the lattice, 
\begin{eqnarray}
T_a(\lambda )=L_{aN}(\lambda )L_{aN-1}(\lambda )\cdots L_{a1}(\lambda ), 
\end{eqnarray}
where the subscript $a$ represents the auxiliary space,
and $1,\cdots, N$ represent the quantum spaces in which
the tensor product is in the graded sense. Explicitely we write\cite{EK}
\begin{eqnarray}
\{ [T(\lambda )]^{ab}\}^{\alpha _1\cdots \alpha _N}
_{\beta _1\cdots \beta _N}
=L_N(\lambda )_{a\alpha _N}^{c_N\beta _N}
L_{N-1}(\lambda )_{c_N\alpha _{N-1}}^{c_{N-1}\beta _{N-1}}
\cdots L_1(\lambda )_{c_2\alpha _1}^{b\beta _1}
(-1)^{\sum _{j=2}^N(\epsilon _{\alpha _j}+\epsilon _{\beta _j})
\sum _{i=1}^{j-1}\epsilon _{\alpha _i}}.
\end{eqnarray}
This definition is different from the non-graded case because 
we have the graded Yang-Baxter equation(\ref{GYBE}).
By repeatedly using the Yang-Baxter relation (\ref{YBR}),
one can prove easily that
the monodromy matrix also satisfies the Yang-Baxter relation,
\begin{eqnarray}
R(\lambda -\mu )T_1(\lambda )T_2(\mu )
=T_2(\mu )T_1(\lambda )R(\lambda -\mu ).
\label{YBR1}
\end{eqnarray}
For periodic boundary condition, the transfer
matrix $\tau _{peri}(\lambda )$
of this model is defined as the supertrace of the
monodromy matrix in the auxiliary space,
\begin{eqnarray}
\tau _{peri}(\lambda )=strT(\lambda )
=\sum (-1)^{\epsilon _a}T(\lambda )_{aa}.
\end{eqnarray}
As a consequence of the Yang-Baxter relation (\ref{YBR1}) and the unitarity
property of the R-matrix, we can prove that
the transfer matrix commutes with each other for different spectral
parameters,
\begin{eqnarray}
[\tau _{peri}(\lambda ),\tau _{peri}(\mu )]=0. 
\end{eqnarray}
In this sense we say that the model is
integrable. Expanding the transfer matrix in the powers of $\lambda $,
we can find conserved operators. And the Hamiltonian is defined as
\begin{eqnarray}
H=sin(\eta )
\frac {d\ln [\tau _{peri}(\lambda )]}{d\lambda }|_{\lambda =0}
=\sum _{j=1}^NH_{j,j+1}=
\sum _{j=1}^NP_{j,j+1}L'_{j,j+1}(0),
\end{eqnarray}
where $P_{ij}$ is the graded permutation operator expressed as
$P_{ac}^{bd}=\delta _{ad}\delta _{bc}(-1)^{\epsilon _a\epsilon _c}$.
The explicit expression of the Hamiltonian has already been
presented in equation (2).

In this paper, we consider the reflecting boundary condition case.
In addition to the Yang-Baxter equation, a reflection equation should
be used in proving the commutativity of the transfer matrix with
boundaries. The reflection equation takes the form \cite{S3},
\begin{eqnarray}
R_{12}(\lambda -\mu )K_1(\lambda )R_{21}(\lambda +\mu )
K_2(\mu )=K_2(\mu )R_{12}(\lambda +\mu )K_1(\lambda )         
R_{21}(\lambda -\mu ).
\end{eqnarray}
For the graded case, the reflection equation remains the
same as the above.
We only need to change the usual tensor product to the graded
tensor product. We write it explicitly as
\begin{eqnarray}
&&R(\lambda -\mu )_{a_1a_2}^{b_1b_2}K(\lambda )_{b_1}^{c_1}
R(\lambda +\mu )_{b_2c_1}^{c_2d_1}                           
K(\mu )_{c_2}^{d_2}(-)^{(\epsilon _{b_1}+\epsilon _{c_1})\epsilon _{b_2}}
\nonumber \\
&=&K(\mu )_{a_2}^{b_2}R(\lambda +\mu )_{a_1b_2}^{b_1c_2}     
K(\lambda )_{b_1}^{c_1}
R(\lambda -\mu )_{c_2c_1}^{d_2d_1}                            
(-)^{(\epsilon _{b_1}+\epsilon _{c_1})\epsilon _{c_2}}.
\label{RK}
\end{eqnarray}
Instead of the monodromy matrix $T(\lambda )$ for periodic boundary
conditions, we consider the double-row monodromy matrix
\begin{eqnarray}
{\cal {T}}(\lambda )=T(\lambda )K(\lambda )T^{-1}(-\lambda )
\end{eqnarray}
for the reflecting boundary conditions. Using the Yang-Baxter relation, and
considering the boundary K-matrix which satisfies
the reflection equation, one
can prove that the double-row monodromy matrix ${\cal {T}}(\lambda )$
also satisfies the reflection equation,
\begin{eqnarray}
&&R(\lambda -\mu )_{a_1a_2}^{b_1b_2}{\cal {T}}(\lambda )_{b_1}^{c_1}
R(\lambda +\mu )_{b_2c_1}^{c_2d_1}                           
{\cal {T}}(\mu )_{c_2}^{d_2}
(-)^{(\epsilon _{b_1}+\epsilon _{c_1})\epsilon _{b_2}}
\nonumber \\
&=&{\cal {T}}(\mu )_{a_2}^{b_2}R(\lambda +\mu )_{a_1b_2}^{b_1c_2}     
{\cal {T}}(\lambda )_{b_1}^{c_1}
R(\lambda -\mu )_{c_2c_1}^{d_2d_1}                            
(-)^{(\epsilon _{b_1}+\epsilon _{c_1})\epsilon _{c_2}}.
\label{RT}
\end{eqnarray}

Next, we study the properties of the R-matrix.
We define the super-transposition $st$ as
\begin{eqnarray}
(A^{st})_{ij}=A_{ji}(-1)^{(\epsilon _i+1)\epsilon _j}.
\end{eqnarray}
We also define the inverse of 
the super-transposition $\bar {st}$ as 
$\{ A^{st}\} ^{\bar {st}}=A$.

One can prove directly that the
R-matrix satisfy the following unitarity and cross-unitarity
relations:
\begin{eqnarray}
&&R_{12}(\lambda )R_{21}(-\lambda )=\rho (\lambda )\cdot id.,~~~~
\rho (\lambda )=sin(\eta +\lambda )sin(\eta -\lambda ),    \\
&&R_{12}^{st_1}(\eta -\lambda )M_1
R_{21}^{st_1}(\lambda )M_1^{-1}={\tilde {\rho }}(\lambda )\cdot id.,~~~~
\tilde {\rho }(\lambda )=sin(\lambda )sin(\eta -\lambda ).
\end{eqnarray}
Here the matrix $M$ is determined by the R-matrix.
For three different gradings, the forms of $M$ are different.
We have $M=diag.(e^{2i\eta },1,1)$ for FFB grading,
$M=diag.(1,1,e^{-2i\eta })$ for BFF grading, and $M=1$ for FBF grading.
We also have a property
\begin{eqnarray}
[R_{12}^{st_1st_2}(\lambda ), M\otimes M]=0.
\end{eqnarray}
The cross-unitarity relation can also be written as follows,
\begin{eqnarray}
\left\{ M_1^{-1}R_{12}^{st_1st_2}(\eta -\lambda )M_1
\right\} ^{\bar {st}_2}R_{21}^{st_1}(\lambda )&=&\tilde {\rho }(\lambda ),\\
R_{12}^{st_1}(\lambda )\left\{ M_1R_{21}^{st_1st_2}(\eta -\lambda )
M_1^{-1}\right\} ^{\bar {st}_2}&=&\tilde {\rho }(\lambda ).
\end{eqnarray}
In order to construct the commuting transfer matrix with boundaries,
besides the
reflection equation, we need the dual reflection equation.
In general, the dual reflection equation which depends on
the unitarity and cross-unitarity relations of the R-matrix   
takes different forms for different models. 
For the models considered in this paper,
we can write the dual
reflection equation in the following form:
\begin{eqnarray}
&&R_{21}^{st_1st_2}(\mu -\lambda )
{K_1^+}^{st_1}(\lambda )M_1^{-1}R_{12}^{st_1st_2}(\eta -\lambda -\mu )
M_1{K_2^+}^{st_2}(\mu )     
\nonumber \\
&=&{K_2^+}^{st_2}(\mu )M_1R_{21}^{st_1st_2}(\eta -\lambda -\mu )
M_1^{-1}{K_1^+}^{st_1}(\lambda )R_{12}^{st_1st_2}(\mu -\lambda ).
\label{DRK}
\end{eqnarray}
Then the transfer matrix with boundaries is defined as
\begin{eqnarray}
t(\lambda )=strK^+(\lambda ){\cal {T}}(\lambda ).
\label{tran}
\end{eqnarray}
The commutativity of $t(\lambda )$ can be proved by using the
unitarity and cross-unitarity relations, the reflection equation and 
the dual reflection equation.
With a normalization $K(0)=id.$, the Hamiltonian can be obtained as
\begin{eqnarray}
H&\equiv &{1\over 2}sin(\eta )\frac {d\ln t(\lambda )}
{d\lambda}|_{\lambda =0}
\nonumber \\
&=&\sum _{j=1}^{N-1}P_{j,j+1}L'_{j,j+1}(0)+
{1\over 2}sin(\eta )K'_1(0)
+\frac {str_aK_a^+(0)P_{Na}L_{Na}'(0)}{str_{a}K_a^+(0)}.
\label{hami}
\end{eqnarray}

\section{Integrable boundary impurities for $t-J$ model with
FFB grading}
Boundary higher spin impurities for the supersymmetric $t-J$ model
was considered in Ref.\cite{WDHP,ZGLG}.
In the previous work\cite{FWY}, we studied boundary higher
spin impurities for the generalized supersymmetric $t-J$
model. 
In the calculation, the FFB grading is used, i.e.
$\epsilon _1=\epsilon _2=1,\epsilon _3=0$.
In order to obtain the Hamiltonian from the transfer
matrix, we use the following representations:
\begin{eqnarray}
S_k=e^k_{21}, ~~~S_k^{\dagger }=e^k_{12},
~~~S_k^z={1\over 2}(e^k_{22}-e^k_{11}),
\label{rep1}
\end{eqnarray}
and
\begin{eqnarray}
&&Q_{k,1}=(1-n_{k,-1})c_{k,1}=e^k_{32},
~~~~Q_{k,1}^{\dagger }=(1-n_{k,-1})c_{k,1}^{\dagger }=e^k_{23},
~~~~Q_{k,-1}=(1-n_{k,1})c_{k,-1}=e^k_{31},
\nonumber \\
&&Q_{k,-1}^{\dagger }=(1-n_{k,1})c_{k,-1}^{\dagger }=e^k_{13},
~~~~T_k=1-{1\over 2}n_k={1\over 2}(e^k_{11}+e^k_{22})+e^k_{33},
\label{rep2}
\end{eqnarray}
where $e_{ij}^k$ is a $3\times 3$ matrix acting on the $k$-th space with
elements $(e_{ij}^k)_{\alpha \beta }=\delta _{i\alpha }
\delta _{j\beta }$. 

We present some results about the boundary higher spin
impurities for the generalized supersymmetric $t-J$ model.
The detailed calculations can be found in Ref.\cite{FWY}.
We suppose that
K-matrix takes the form,
\begin{eqnarray}
K(\lambda )=\left( \begin{array}{ccc}
A(\lambda ) &B(\lambda ) &0\\
C(\lambda ) &D(\lambda ) &0\\
0&0&1\end{array}\right).
\label{K}
\end{eqnarray}
Inserting this matrix into the reflection equation (\ref{RK}),
we find the following solutions,
\begin{eqnarray}
A(\lambda )&=&g(\lambda )\left( e^{-4i\lambda }sin(\lambda +c-s\eta )
sin(\lambda +c+\eta +s\eta )-sin(2\lambda )sin(\lambda +c-{\bf S^z}\eta )
e^{-i(3\lambda +c+\eta +{\bf S^z}\eta )}\right),
\nonumber \\
B(\lambda )&=&g(\lambda )sin(\eta )sin(2\lambda )
e^{-i(2\lambda -c+{\bf S^z}\eta )}{\bf S^-},
\nonumber \\
C(\lambda )&=&g(\lambda )sin(\eta )sin(2\lambda )
e^{-i(2\lambda +c+{\bf S^z}\eta )}{\bf S^+},
\nonumber \\
D(\lambda )&=&g(\lambda )\left( sin(\lambda +c-s\eta )
sin(\lambda +c+\eta +s\eta )-sin(2\lambda )sin(\lambda +c+
{\bf S^z}\eta )e^{-i(\lambda -c-\eta +{\bf S^z}\eta )}\right),
\label{k}
\end{eqnarray}
where $g(\lambda )=1/sin(\lambda -c-\eta -s\eta )sin(\lambda -c+s\eta )$.
${\bf S^z, S}$ and ${\bf S^{\dagger }}$ are
spin-$s$ operators satisfying the following commutation relations,
\begin{eqnarray}                                           
[{\bf S^z}, {\bf S^{\pm }}]=\pm {\bf S^{\pm }},
~~~~~~~[{\bf S^+,S^-}]=\frac {sin(2{\bf S^z}\eta )}{sin(\eta )}.
\end{eqnarray}
We suppose $K^+$ has
the similar form as $K$. By direct calculation, we can find
$R^{st_1st_2}_{12}(\lambda )=I_1R_{21}(\lambda )I_1$ with $I=diag.(-1,-1,1)$.
For the form (\ref{K}), we have $IK(\lambda )I=K(\lambda )$. Then with
the help of property $[M_1M_2, R(\lambda )]=0$, 
we see that there is an isomorphism between $K$ and $K^+$:
\begin{eqnarray}
K(\lambda ):\rightarrow {K^+}^{st}(\lambda )=K({\eta \over 2}-\lambda )M.
\label {ISOM}
\end{eqnarray}
Given a solution to the reflection equation (\ref{RK}), we can also
find a solution to the dual reflection equation (\ref{DRK}).
Note that in the
sense of the transfer matrix, the reflection equation and the dual
reflection equation are independent of each other.
For other gradings, BFF and FBF, the isomorphism (\ref{ISOM})
does not hold.

By definition in equation (\ref{hami}), and using the explicit form
of the boundary reflecting matrices $K$ and $K^+$,
we can find the boundary impurity terms. 
The boundary impurity coupled to site 1 is written as
\begin{eqnarray}
H_1&=&\frac {2}{sin(c+\eta +s\eta )sin(c-s\eta )}
e^{-i{\bf S^z}\eta }
[e^{ic}{\bf S^-}S_1^{\dagger }+e^{-ic}{\bf S^+}S_1
\nonumber \\
&&+(e^{-i(c+\eta )}sin(c-{\bf S^z}\eta )S_1^z
-e^{i(c+\eta )}sin(c+{\bf S^z}\eta )S_1^z)
\nonumber \\
&&+(e^{-i(c+\eta )}sin(c-{\bf S^z}\eta )
+e^{i(c+\eta )}sin(c+{\bf S^z}\eta ))T_1]
+4i(T_1+S_1^z).
\label{imp}
\end{eqnarray}
The impurity coupled to site N is in a similar form.
We remark here that in the rational limit and by some redefinition,
(\ref{imp}) becomes the usual spin-exchange term
${\bf S\cdot S_1}$ between the impurity spin and the spin in site 1.

By using the algebraic Bethe ansatz method, we can find the eigenvalue
of the transfer matrix and the Bethe ansatz equations. Here we just list
the results. The energy spectrum of the Hamiltonian
(\ref{hami}) is given by
\begin{eqnarray}
E&=&(N-2)cos(\eta )+\sum _{i=1}^n
\frac {sin^2(\eta )}{sin(\mu _i)sin(\mu _i+\eta )}
\nonumber \\
&&-sin^2(\eta )
\left[ \frac {1}{sin(\tilde {c}-\eta +\tilde {s}\eta )
sin(\tilde {c}+\tilde {s}\eta )}
+\frac {1}{sin(\tilde {c}-\eta -\tilde {s}\eta )
sin(\tilde {c}-2\eta -\tilde {s}\eta )}
\right] ,
\end{eqnarray}
where $\mu _1, \cdots ,\mu _n$ and $\mu ^{(1)}_1, \cdots ,\mu ^{(1)}_m$
should satisfy the Bethe ansatz equations
\begin{eqnarray}
&&
\frac {sin(\mu _j^{(1)}+c+\eta +s\eta )sin(\mu _j^{(1)}-c-\eta +s\eta )
sin(\mu _j^{(1)}+\tilde {c}-\eta +\tilde {s}\eta )
sin(\mu _j^{(1)}-\tilde {c}-\tilde {s}\eta )               
}
{sin(\mu _j^{(1)}-c-\eta -s\eta )sin(\mu _j^{(1)}+c+\eta -s\eta )
sin(\mu _j^{(1)}+\tilde {c}-\eta -\tilde {s}\eta )
sin(\mu _j^{(1)}-\tilde {c}+\eta -\tilde {s}\eta )
}
\nonumber \\
&&
=
\prod _{i=1}^n
\frac {sin(\mu _j^{(1)}+\mu _i)
sin(\mu _j^{(1)}-\mu _i-\eta )} 
{sin(\mu _j^{(1)}+\mu _i+\eta )
sin(\mu _j^{(1)}-\mu _i)}
\prod _{l=1, \not=j}^m
\frac {sin(\mu _j^{(1)}-\mu ^{(1)}_l+\eta )
sin(\mu _j^{(1)}+\mu ^{(1)}_l+\eta )}
{sin(\mu _j^{(1)}-\mu ^{(1)}_l-\eta )
sin(\mu _j^{(1)}+\mu ^{(1)}_l-\eta )},
\nonumber \\
&&\hskip 10truecm~~~~j=1, \cdots , m,
\end{eqnarray}
and 
\begin{eqnarray}
&&\frac 
{sin(\mu _j+\tilde {c}-\eta -\tilde {s}\eta )
sin(\lambda +c+\eta -s\eta )
}
{sin(\mu _j-\tilde {c}+2\eta +\tilde {s}\eta )
sin(\lambda -c+s\eta )}
=
\frac {sin^{2N}(\mu _j+\eta )}{sin^{2N}(\mu _j)}
\prod _{l=1}^m
\frac {sin(\mu _j-\mu _l^{(1)})
sin(\mu _j+\mu _l^{(1)})}
{sin(\mu _j-\mu _l^{(1)}+\eta )
sin(\mu _j+\mu _l^{(1)}+\eta )},
\nonumber \\
&&\hskip 10truecm~~~~j=1,\cdots ,n.
\end{eqnarray}

\section{Boundary impurities for the case of BFF grading}
We have $\epsilon _1=0, \epsilon _2=\epsilon _3=1$ for BFF grading.
For a supermatrix $X$, the Grassmann parities for its entries $X_{ij}$
are defined as $\epsilon _i+\epsilon _j$.
We can change the representations (\ref{rep1},\ref{rep2}) in the last
section to satisfy the grading by performing a change
from 1,2,3 to 2,3,1. Explicitly, we have
\begin{eqnarray}
&&S_k=e^k_{32}, ~~~S_k^{\dagger }=e^k_{23},
~~~S_k^z={1\over 2}(e^k_{33}-e^k_{22}),
\nonumber \\
&&Q_{k,1}=(1-n_{k,-1})c_{k,1}=e^k_{13},
~~~~Q_{k,1}^{\dagger }=(1-n_{k,-1})c_{k,1}^{\dagger }=e^k_{31},
~~~~Q_{k,-1}=(1-n_{k,1})c_{k,-1}=e^k_{12},
\nonumber \\
&&Q_{k,-1}^{\dagger }=(1-n_{k,1})c_{k,-1}^{\dagger }=e^k_{21},
~~~~T_k=1-{1\over 2}n_k={1\over 2}(e^k_{22}+e^k_{33})+e^k_{11}.
\end{eqnarray}

For the nested algebraic Bethe ansatz method for BFF grading,
the low level $r$-matrix is BF grading which is different
from the FF grading $r$-matrix in the case of FFB grading, because
the graded calculation for FF grading $r$-matrix
is actually equal to the non-graded case.

\subsection{Solutions to the reflection equation and the dual
reflection equation}
We begin with the explicit form of the R-matrix,
\begin{eqnarray}
R(\lambda )=\left(
\begin{array}{ccccccccc}
w(\lambda )&0&0&0&0&0&0&0&0\\                                        
0&b(\lambda )&0&c_-(\lambda )&0&0&0&0&0\\
0&0&b(\lambda )&0&0&0&c_-(\lambda )&0&0\\
0&c_+(\lambda )&0&b(\lambda )&0&0&0&0&0\\
0&0&0&0&a(\lambda )&0&0&0&0\\
0&0&0&0&0&b(\lambda )&0&-c_-(\lambda )&0\\
0&0&c_+(\lambda )&0&0&0&b(\lambda )&0&0\\
0&0&0&0&0&-c_+(\lambda )&0&b(\lambda )&0\\
0&0&0&0&0&0&0&0&a(\lambda )
\end{array}\right),
\end{eqnarray}
where we use the notations,
\begin{eqnarray}
a(\lambda )=sin(\lambda -\eta ), ~w(\lambda )=sin(\lambda +\eta ),
~b(\lambda )=sin(\lambda ),~c_{\pm }(\lambda )=e^{\pm i\lambda }sin(\eta ).
\end{eqnarray}
We still assume that the reflecting K-matrix operator takes the
form
\begin{eqnarray}
K(\lambda )=\left( \begin{array}{ccc}
A(\lambda ) &B(\lambda ) &0\\
C(\lambda ) &D(\lambda ) &0\\
0&0&1\end{array}\right).
\label{K1}
\end{eqnarray}
Inserting this matrix into the reflection equation (\ref{RK}), we can find
the following non-trivial relations:
\begin{eqnarray}
&&r(\lambda -\mu )_{a_1a_2}^{b_1b_2}K(\lambda )_{b_1}^{c_1}
r(\lambda +\mu )_{b_2c_1}^{c_2d_1}                           
K(\mu )_{c_2}^{d_2}(-1)^{(\epsilon _{b_1}+\epsilon _{c_1})\epsilon _{b_2}}
\nonumber \\
&=&K(\mu )_{a_2}^{b_2}r(\lambda +\mu )_{a_1b_2}^{b_1c_2}
K(\lambda )_{b_1}^{c_1}
r(\lambda -\mu )_{c_2c_1}^{d_2d_1}
(-1)^{(\epsilon _{b_1}+\epsilon _{c_1})\epsilon _{c_2}},
\label{rk}
\end{eqnarray}
and
\begin{eqnarray}
K(\lambda )_{a_1}^{b_1}K(\mu )_{b_1}^{d_1}=
K(\mu )_{a_1}^{b_1}K(\lambda )_{b_1}^{d_1},
\label{rk1}
\end{eqnarray}
\begin{eqnarray}
\delta _{a_1d_1}sin(\lambda -\mu )e^{-i(\lambda +\mu )}
+sin(\lambda +\mu )e^{i(\lambda -\mu )}K(\lambda )_{a_1}^{d_1}
\nonumber \\
=e^{-i(\lambda -\mu )}sin(\lambda +\mu )K(\mu )_{a_1}^{d_1}  
+e^{i(\lambda +\mu )}K(\mu )_{a_1}^{b_1}K(\lambda )_{b_1}^{d_1},
\label{rk2}
\end{eqnarray}
where all indices take values 1,2,
and we have the BF grading, i.e. $\epsilon _1=0, \epsilon _2=1$.
We also have introduced the notation
\begin{eqnarray}
r_{12}(\lambda )=\left( \begin{array}{cccc}
sin(\lambda +\eta )&0&0&0\\
0&sin(\lambda )&sin(\eta )e^{-i\lambda }&0\\
0&sin(\eta )e^{i\lambda }&sin(\lambda )&0\\                 
0&0&0&sin(\lambda -\eta )\end{array}\right).
\label{r}
\end{eqnarray}
This matrix $r(\lambda )$ has the BF grading and satisfies
the graded Yang-Baxter equation. To find a solution to
these relations, we first construct a solution to (\ref{rk})
by $r$-matrix (\ref{r}) in the same way as a construction of
a double-row monodromy matrix. Then we check other relations
(\ref{rk1},\ref{rk2}). After some tedious calculations,
we finally obtain the following results
\begin{eqnarray}
A(\lambda )&=&{1\over 2}
g(\lambda )\left[ sin(\lambda +c+\eta )
sin(\lambda -c+\eta )+sin(\lambda +c)sin(\lambda -c)
\right. \nonumber \\
&&\left.
+sin^2(\eta )e^{-2i\lambda }+sin({\eta })sin(2\lambda )e^{i\eta }\sigma ^z
\right] ,
\nonumber \\
B(\lambda )&=&g(\lambda )sin(\eta )sin(2\lambda )
\sigma ^-,
\nonumber \\
C(\lambda )&=&g(\lambda )sin(\eta )sin(2\lambda )
\sigma ^+,
\nonumber \\
D(\lambda )&=&{1\over 2}g(\lambda )
\left[ sin(\lambda +c-\eta )
sin(\lambda -c-\eta )+sin(\lambda +c)sin(\lambda -c)
\right. \nonumber \\
&&\left. +sin^2(\eta )e^{2i\lambda }
+sin(\eta )sin(2\lambda )e^{i\eta }\sigma ^z\right],
\label{k1}
\end{eqnarray}
where
\begin{eqnarray}
g(\lambda )=\frac {-e^{-2i\lambda }}
{sin(\lambda -c+\eta )sin(\lambda -c-\eta )},
\end{eqnarray}
$\sigma ^{\pm }={1\over 2}(\sigma ^x\pm i\sigma ^y)$,
$\sigma ^x,\sigma ^y,\sigma ^z$ are Pauli matrices,
and $c$ is an arbitrary parameter. In the graded method,
reflecting K-matrix (\ref{K1}) is a supermatrix with
BFF grading. That means $B(\lambda )$ and $C(\lambda )$
are Grassmann odd, and $A(\lambda )$ and $D(\lambda )$
are Grassmann even. Therefore, $\sigma ^{\pm }$ are
Grassmann odd and can be represented by fermion operators
$a_L$ and $a^{\dagger }_L$,
$\sigma ^+=a_L, \sigma ^-=a^{\dagger }_L$.
And $\sigma ^z$ is Grassmann even,
and can be represented as $\sigma ^z=1-2n_L$,
where we denote $n_L=a^{\dagger }_La_L$.

Next, let us solve the dual reflection equation (\ref{DRK})
for BFF grading. For FFB grading in section 3, we have
an isomorphism between $K$ and $K^+$ (\ref{ISOM}), this
is not the case here.
Considering the form of $K$ (\ref{K1})
we have for BFF grading, and $R_{12}^{st_1st_2}(\lambda )
=I_1R_{21}(\lambda )I_1$, where $I=diag.(1, -1, -1)$,
we do not have the relation $IK(\lambda )I=K(\lambda )$ now.
Due to this fact, we have to solve the dual
reflection equation (\ref{DRK}) independently. The strategy is almost
the same as that for reflection equation. Here we just list the
final results. We assume $K^+$ takes the form
\begin{eqnarray}
K^+(\lambda )=\left( \begin{array}{ccc}
A^+(\lambda ) &B^+(\lambda ) &0\\
C^+(\lambda ) &D^+(\lambda ) &0\\
0&0&1\end{array}\right).
\label{dualK}
\end{eqnarray}
After some tedious calculations, we have
\begin{eqnarray}
A^+(\lambda )&=&{1\over 2}
g^+(\lambda )\left[ sin(\lambda +\tilde {c}+{\eta \over 2})
sin(\lambda -\tilde {c}+{\eta \over 2})
+sin(\lambda +\tilde {c}-{\eta \over 2})
sin(\lambda -\tilde {c}-{\eta \over 2})
\right. \nonumber \\
&&\left.
-sin^2(\eta )e^{i(2\lambda -\eta )}
+sin({\eta })sin(2\lambda -\eta )e^{-i\eta }\tilde {\sigma }^z
\right] ,
\nonumber \\
B^+(\lambda )&=&g^+(\lambda )e^{-i\eta }sin(\eta )sin(2\lambda -\eta )
\tilde {\sigma }^-,
\nonumber \\
C^+(\lambda )&=&g^+(\lambda )e^{-i\eta }sin(\eta )sin(2\lambda -\eta )
\tilde {\sigma }^+,
\nonumber \\
D^+(\lambda )&=&{1\over 2}g^+(\lambda )
\left[ sin(\lambda +\tilde {c}-{3\eta \over 2})
sin(\lambda -\tilde {c}-{3\eta \over 2})
+sin(\lambda +\tilde {c}-{\eta \over 2})
sin(\lambda -\tilde {c}-{\eta \over 2})
\right. \nonumber \\
&&\left. -sin^2(\eta )e^{-i(2\lambda -\eta )}
+sin(\eta )sin(2\lambda -\eta )e^{-i\eta }
\tilde {\sigma }^z\right],
\end{eqnarray}
where
$\tilde {c}$ is an arbitrary parameter and
\begin{eqnarray}
g^+(\lambda )=\frac {-e^{i(2\lambda +\eta )}}
{sin^2(\lambda -\tilde {c}-{\eta \over 2})}.
\end{eqnarray}
This reflecting K-matrix
$K^+$ is also a supermatrix with BFF grading. Therefore,
$B^+(\lambda )$ and $C^+(\lambda )$ are Grassmann odd,
$A^+(\lambda ),D^+(\lambda )$ are Grassmann even.
Here we use the representation
$\tilde {\sigma }^+=a_R,
\tilde {\sigma }^-=a^{\dagger }_R, \tilde {\sigma }^z=1-2n_R$,
and $n_R=a^{\dagger }_Ra_R$.

We can thus obtain the boundary impurity terms of the Hamiltonian
defined by the reflecting matrices,
\begin{eqnarray}
H_1&=&\frac {2sin(\eta )}{sin(\eta +c)sin(\eta -c)}
(1-n_{1,1})[a_L^{\dagger }c_{1,-1}
+c_{1,-1}^{\dagger }a_L+{\frac {e^{i\eta }}{2}}(1-2n_L)]
\nonumber \\
&&+(1-n_{1,1})
\left( \frac {2sin(2c)}{sin(\eta +c)sin(c-\eta )}-4i\right)
+\frac {sin(\eta )e^{-i\eta }}{sin(\eta +c)sin(\eta -c)}
(3T_1+S_1^z-2).
\end{eqnarray}
$H_N$ has a similar form.

\subsection{Algebraic Bethe ansatz method}
We shall use the nested algebraic Bethe ansatz method to
obtain the eigenvalue of the transfer matrix with open boundary
conditions.
We denote the double-row monodromy matrix as
\begin{eqnarray}
{\cal {T}}(\lambda )
&=&\left( \begin{array}{ccc}
{\cal {A}}_{11}(\lambda )&{\cal {A}}_{12}(\lambda ) &
{\cal {B}}_1(\lambda )\\
{\cal {A}}_{21}(\lambda )&{\cal {A}}_{22}(\lambda )
&{\cal {B}}_2(\lambda )\\
{\cal {C}}_1(\lambda )&{\cal {C}}_2(\lambda )&{\cal {D}}
(\lambda )\end{array}
\right).
\end{eqnarray}
For convenience, we introduce the following transformations
\begin{eqnarray}
{\cal {A}}_{ab}(\lambda )=\tilde {\cal {A}}_{ab}(\lambda )
-\delta _{ab}
\frac {e^{-2i\lambda }sin(\eta )}{sin(2\lambda -\eta )}
{\cal {D}}(\lambda ).
\end{eqnarray}
Note that this transfermation is different from that of FFB grading.
The transfer matrix (\ref{tran}) can be rewritten as
\begin{eqnarray}
t(\lambda )&=&A^+(\lambda ){\cal {A}}_{11}(\lambda )
-D^+(\lambda ){\cal {A}}_{22}(\lambda )-{\cal {D}}(\lambda )
\nonumber \\
&=&A^+(\lambda )\tilde {\cal {A}}_{11}(\lambda )
-D^+(\lambda )\tilde {\cal {A}}_{22}(\lambda )
-U_3^+(\lambda ){\cal {D}}(\lambda ),
\end{eqnarray}
where
\begin{eqnarray}
U_3^+(\lambda )=
1+\frac {e^{-2i\lambda }sin(\eta )}{sin(2\lambda -\eta )}
[A^+(\lambda )-D^+(\lambda )].
\end{eqnarray}
We define a reference state in the $n$-th quantum space
as $|0>_n=(0, 0, 1)^t$, and reference states for the
boundary opeators as $\sigma ^-|0>_L=0, \sigma ^z|0>_L=-|0>_L,
\sigma ^+|0>_L\not= 0$, and
$\tilde {\sigma }^-|0>_R=0, \tilde {\sigma }^z|0>_R=-|0>_R,
\tilde {\sigma }^+|0>_R\not= 0$,
where L and R stand for the left and right boundary sites.
The vacuum state is then defined as
$|0>=|0>_L\otimes_{k=1}^N|0>_k\otimes |0>_R$. Acting the double-row
monodromy matrix on this vacuum state, with the help of
Yang-Baxter relation, we have
\begin{eqnarray}                         
{\cal {B}}_a(\lambda )|0>&=&0,\nonumber \\
{\cal {C}}_a(\lambda )|0>&\not= &0,\nonumber \\
{\cal {D}}(\lambda )|0>&=&sin^{2N}(\lambda -\eta )|0>,\nonumber \\
\tilde {\cal {A}}_{ab}(\lambda )|0>&=&sin^{2N}(\lambda )
[K(\lambda )_a^b+
\delta _{ab}\frac {sin(\eta )e^{-2i\lambda }}{sin(2\lambda -\eta )}]
|0>
=W_{ab}(\lambda )sin^{2N}(\lambda )|0>,
\end{eqnarray}
where
\begin{eqnarray}
&&W_{12}(\lambda )=0, \nonumber \\
&&W_{11}(\lambda )=\frac {-e^{-2i\lambda }sin(2\lambda )}
{sin(2\lambda -\eta )sin(\lambda -c+\eta )
sin(\lambda -c-\eta )}
[sin(\lambda +c-\eta )
sin(\lambda -c)
\nonumber \\
&&~~~~~+sin^2(\eta )e^{-i(2\lambda -\eta )}]
\nonumber \\
&&W_{22}(\lambda )=-e^{-2i\lambda }\frac {sin(2\lambda )
sin(\lambda +c-2\eta )}{sin(2\lambda -\eta )sin(\lambda -c+\eta )}.
\label{W}
\end{eqnarray}

As mentioned in section 2, the double-row monodromy matrix satisfies
the reflection equation (\ref{RT}). We thus have the following
commutation relations:
\begin{eqnarray}
{\cal {C}}_{d_1}(\lambda ){\cal {C}}_{d_2}(\mu )
&=&\frac {r_{12}(\lambda -\mu )_{c_2c_1}^{d_2d_1}}{sin(\lambda -\mu -\eta )}
(-1)^{1+\epsilon _{d_1}+\epsilon _{c_2}+
\epsilon _{c_1}\epsilon _{c_2}}
{\cal {C}}_{c_2}(\mu ){\cal {C}}_{c_1}(\lambda ),
\end{eqnarray}

\begin{eqnarray}
{\cal {D}}(\lambda ){\cal {C}}_{d}(\mu )
&=&\frac {sin(\lambda +\mu )sin(\lambda -\mu +\eta )}
{sin(\lambda +\mu -\eta )sin(\lambda -\mu )}
{\cal {C}}_{d}(\mu )
{\cal {D}}(\lambda )
\nonumber \\
&&-\frac {sin(2\mu )sin(\eta )e^{i(\lambda -\mu )}}
{sin(\lambda -\mu )sin(2\mu -\eta )}  
{\cal {C}}_{d}(\lambda )
{\cal {D}}(\mu )
+\frac {sin(\eta )e^{i(\lambda +\mu )}}
{sin(\lambda +\mu -\eta )}
{\cal {C}}_b(\lambda )\tilde {\cal {A}}_{bd}(\mu ),
\end{eqnarray}

\begin{eqnarray}
\tilde {\cal {A}}_{a_1d_1}(\lambda ){\cal {C}}_{d_2}(\mu )
&=&(-1)^{\epsilon _{a_1}+\epsilon _{d_1}+\epsilon _{c_1}\epsilon _{b_2}
+\epsilon _{d_1}\epsilon _{d_2}}
\frac {r_{12}(\lambda +\mu -\eta )_{a_1c_2}^{c_1b_2}
r_{21}(\lambda -\mu )_{b_1b_2}^{d_1d_2}}
{sin(\lambda +\mu -\eta )sin(\lambda -\mu )}                
{\cal {C}}_{c_2}(\mu )
\tilde {\cal {A}}_{c_1b_1}(\lambda )
\nonumber \\
&&-(-1)^{\epsilon _{a_1}(1+\epsilon _{b_1})+\epsilon _{d_1}}
\frac {sin(\eta )e^{-i(\lambda -\mu )}}
{sin(\lambda -\mu )sin(2\lambda -\eta )}  
r_{12}(2\lambda -\eta )_{a_1b_1}^{b_2d_1}
{\cal {C}}_{b_1}(\lambda )
\tilde {\cal {A}}_{b_2d_2}(\mu )
\nonumber \\
&&+(-1)^{\epsilon _{d_1}+\epsilon _{a_1}(\epsilon _{d_1}+\epsilon _{d_2})}
\frac {sin(2\mu )sin(\eta )e^{-i(\lambda +\mu )}}
{sin(\lambda +\mu -\eta )sin(2\lambda -\eta )sin(2\mu -\eta )}
\nonumber \\
&&~~~~~\times r_{12}(2\lambda +\eta )_{a_1b_2}^{d_2d_1}
{\cal {C}}_{b_2}(\lambda ){\cal {D}}(\mu ).
\end{eqnarray}
Here the indices take values 1,2, and the Grassmann parities
are BF, $\epsilon _1=0, \epsilon _2=1$. 

By use of the standard algebraic Bethe ansatz method, acting the above
defined transfer matrix on the ansatz of eigenvector
${\cal {C}}_{d_1}(\mu _1)
{\cal {C}}_{d_2}(\mu _2)\cdots {\cal {C}}_{d_n}(\mu _n)|0>
F^{d_1\cdots d_n}$, we have
\begin{eqnarray}
&&t(\lambda )
{\cal {C}}_{d_1}(\mu _1)
{\cal {C}}_{d_2}(\mu _2)\cdots {\cal {C}}_{d_n}(\mu _n)|0>F^{d_1\cdots d_n}
\nonumber \\
&=&-U_3^+(\lambda )sin^{2N}(\lambda -\eta )
\prod _{i=1}^n
\frac {sin(\lambda +\mu _i)sin(\lambda -\mu _i+\eta )}
{sin(\lambda +\mu _i-\eta )sin(\lambda -\mu _i)}
{\cal {C}}_{d_1}(\mu _1)
\cdots {\cal {C}}_{d_n}(\mu _n)|0>F^{d_1\cdots d_n}
\nonumber \\
&+&sin^{2N}(\lambda )                                        
\prod _{i=1}^n
\frac {1}
{sin(\lambda -\mu _i)sin(\lambda +\mu _i-\eta )}
{\cal {C}}_{c_1}(\mu _1)
\cdots {\cal {C}}_{c_n}(\mu _n)|0>
t^{(1)}(\lambda )^{c_1\cdots c_n}_{d_1\cdots d_n}            
F^{d_1\cdots d_n}
\nonumber \\
&+&u.t.,
\end{eqnarray}
where
\begin{eqnarray}
U_3^+=\frac {sin(\lambda -\tilde {c}+{\eta \over 2})
sin(\lambda -\tilde {c}-{3\eta \over 2})}
{sin^2(\lambda -\tilde {c}-{\eta \over 2})}.
\end{eqnarray}
The nested transfer matrix
$t^{(1)}(\lambda )$ is defined as
\begin{eqnarray}
t^{(1)}(\lambda )^{c_1\cdots c_n}_{d_1\cdots d_n}
&=&(-)^{\epsilon _{b}}K^+(\lambda )_b^a
\left\{ r(\lambda +\mu _1-\eta )_{ac_1}^{a_1e_1}
r(\lambda +\mu _2-\eta )_{a_1c_2}^{a_2e_2}\cdots
r(\lambda +\mu _1-\eta )_{a_{n-1}c_n}^{a_ne_n}\right\}
\nonumber \\
&&W_{a_nb_n}(\lambda )
\left\{ r_{21}(\lambda -\mu _n)_{b_ne_n}^{b_{n-1}d_n}
\cdots
r_{21}(\lambda -\mu _2)_{b_2e_2}^{b_1d_2}
r_{21}(\lambda -\mu _1)_{b_1e_1}^{bd_1}
\right\} 
\nonumber \\
&&\times 
(-1)^{\sum _{i=1}^n(\epsilon _{a_i}+\epsilon _{b_i})(1+\epsilon _{e_i})}.
\end{eqnarray}
Thus this nested transfer matrix can still be interpreted as
a transfer matrix with reflecting boundary conditions corresponding
to the anisotropic case\cite{FWW},
\begin{eqnarray}
t^{(1)}(\lambda )=str{K^{(1)}}^+(\lambda ')
T^{(1)}(\lambda ', \{ \mu '_i\} )
K^{(1)}(\lambda ')
{T^{(1)}}^{-1}(-\lambda ', 
\{ \mu '_i\} ),
\label{nest}
\end{eqnarray}                             
where we denote
$x'=x-{\eta \over 2}, x=\lambda ,\mu ,c ,\tilde {c}$.
Notice that this definition is different from
that of the FFB case.
The row-to-row monodromy matrix
$T^{(1)}(\lambda ', \{ \mu '_i\} )$ and
${T^{(1)}}^{-1}(-\lambda ', 
\{ \mu '_i\} )$ are defined respectively as
\begin{eqnarray}                       
T^{(1)}_{aa_n}(\lambda ', 
\{ \mu '_i\} )_{c_1\cdots c_n}^{e_1\cdots e_n}
&=&
r(\lambda '+\mu '_1)_{ac_1}^{a_1e_1}
r(\lambda '+\mu '_2)_{a_1c_2}^{a_2e_2}\cdots
\nonumber \\
&&r(\lambda '+\mu '_1)_{a_{n-1}c_n}^{a_ne_n}
(-1)^{\sum _{j=2}^n(\epsilon _{c_i}+\epsilon _{e_i})
\sum _{i=1}^{j-1}(1+\epsilon _{e_i})},
\label{rowt}
\end{eqnarray}
\begin{eqnarray}
{T^{(1)}}^{-1}_{b_na}(-\lambda ',
\{ \mu '_i\} )_{e_1\cdots e_n}^{d_1\cdots d_n}
&=&r_{21}(\lambda '-\mu '_n)_{b_ne_n}^{b_{n-1}d_n}
\cdots
r_{21}(\lambda '-\mu '_2)_{b_2e_2}^{b_1d_2}
\nonumber \\
&&~~~~r_{21}(\lambda '-\mu '_1)_{b_1e_1}^{ad_1}
(-1)^{\sum _{j=2}^n(\epsilon _{d_i}+\epsilon _{e_i})
\sum _{i=1}^{j-1}(1+\epsilon _{e_i})},
\label{rowt1}
\end{eqnarray}
where we have used the unitarity relation of the r-matrix 
$r_{12}(\lambda )r_{21}(-\lambda )=sin(\eta -\lambda)sin(\eta +\lambda )
\cdot id.$.
We find that the super-tensor product in the above defined
monodromy matrix differs from the original definition.
Nevertheless, as in the periodic boundary condition case, we can
define another graded tensor product as follows \cite{EK}:
\begin{eqnarray}
{F\bar {\otimes }G}_{ac}^{bd}=F_a^bG_c^d(-1)^{(\epsilon _a+\epsilon _b)
(1+\epsilon _c)}.
\label{newgrad}
\end{eqnarray}
Effectively this graded tensor product switches even and odd Grassmann
parities. The graded tensor product in the above monodromy matrices
follows the newly defined rule.

Now, we shall find the eigenvalue of the reduced transfer matrix
(\ref{nest}). Like the original one, we can still
use the graded algebraic Bethe ansatz method. However, we should
be careful about some technical points. The graded tensor
product in the row-to-row monodromy
matrices (\ref{rowt},\ref{rowt1}) are defined by a new definition.
We thus should use a newly defined graded Yang-Baxter relation,
reflection equation and the dual reflection equation.
We should also prove that reflecting matrices appeared in
(\ref{nest}) indeed satisfy their correpsonding reflection
equations.

Define another $r$-matrix\cite{FWW} by
\begin{eqnarray}
\hat{r}(\lambda )_{ac}^{bd}=(-1)^{\epsilon _a+\epsilon _b}
r(\lambda )_{ac}^{bd}.
\label{newr}
\end{eqnarray}
This $r$-matrix has the cross-unitarity,
relation
\begin{eqnarray}
\hat {r}^{st_1}_{12}(-\lambda )
\hat {r}^{st_1}_{21}(\lambda )=-sin^2(\lambda )\cdot id.
\label{newcross}
\end{eqnarray}
With the help of the original Yang-Baxter relation,
we have the following graded Yang-Baxter relation for
the row-to-row monodromy matrix (\ref{rowt}),
\begin{eqnarray}
&&\hat {r}(\lambda _1-\lambda _2 )_{a_1a_2}^{b_1b_2}
T^{(1)}(\lambda _1,\{ \mu _i\} )_{b_1}^{c_1}
T^{(1)}(\lambda _2,\{ \mu _i\} )_{b_2}^{c_2}
(-1)^{(\epsilon _{b_1}+\epsilon _{c_1})
(1+\epsilon _{b_2})}
\nonumber \\
&=&T^{(1)}(\lambda _2,\{ \mu _i\} )_{a_2}^{b_2}
T^{(1)}(\lambda ,\{ \mu _i\} )_{a_1}^{b_1}
\hat {r}(\lambda _1-\lambda _2)_{b_1b_2}^{c_1c_2}
(-1)^{(\epsilon _{a_1}+\epsilon _{b_1})(1+\epsilon _{b_2})}.
\end{eqnarray}
The reflection equation should take the following form,
\begin{eqnarray}
&&\hat {r}(\lambda -\mu )_{a_1a_2}^{b_1b_2}
K^{(1)}(\lambda )_{b_1}^{c_1}
\hat {r}(\lambda +\mu )_{b_2c_1}^{c_2d_1}                           
K^{(1)}(\mu )_{c_2}^{d_2}
(-)^{(\epsilon _{b_1}+\epsilon _{c_1})(1+\epsilon _{b_2})}
\nonumber \\
&=&K^{(1)}(\mu )_{a_2}^{b_2}\hat {r}(\lambda +\mu )_{a_1b_2}^{b_1c_2}     
K^{(1)}(\lambda )_{b_1}^{c_1}
\hat {r}(\lambda -\mu )_{c_2c_1}^{d_2d_1}                            
(-)^{(\epsilon _{b_1}+\epsilon _{c_1})(1+\epsilon _{c_2})}.
\label{newrk}
\end{eqnarray}
Here we find that r-matrix is the newly defined r-matrix $\hat {r}$
(\ref{newr}), and the super tensor-product is also in newly defined
form (\ref{newgrad}). However, the reflecting K-matrix does not
need to be changed, i.e., this reflecting K-matrix also satisfy
the reflection equation (\ref{rk}). We know
the reflecting K-matrix defined in (\ref{nest}) is written as
\begin{eqnarray}
K^{(1)}(\lambda ')&=&e^{-i\eta }\frac {sin(2\lambda '+\eta )}
{sin(2\lambda ')}\left( \begin{array}{cc}
A(\lambda ',c') &B(\lambda ',c')\\
C(\lambda ',c') &D(\lambda ',c')
\end{array}\right).
\label{newk}
\end{eqnarray}
We can easily find that this K-matrix satisfy the relation (\ref{rk}).
Thus we know that it is also a solution to the new reflection
equation (\ref{newrk}).
Similarly, we can deal with the dual reflection equation. 
The reflecting K-matrix defined in (\ref{nest}) 
\begin{eqnarray}
{K^{(1)}}^+(\lambda ')&=&\left( \begin{array}{cc}
A^+(\lambda '+{\eta \over 2})&B^+(\lambda '+{\eta \over 2})\\
C^+(\lambda '+{\eta \over 2})&D^+(\lambda '+{\eta \over 2})
\end{array}\right)
\label{newdualk}
\end{eqnarray}
is shown to satisfy the corresponding dual reflection equation which
is consistent with the cross-unitarity relation (\ref{newcross}).
We thus prove that the nested transfer matrix (\ref{nest}) is
indeed a transfer matrix, and we can as previously use the
graded algebraic Bethe
ansatz method to find its eigenvalues and eigenvectors.

\subsection{Nested algebraic Bethe ansatz}
We show that the problem to find the eigenvalue of $t(\lambda )$ is
changed into a simpler problem to find the eigenvalue of $t^{(1)}(\lambda )$.
We shall still use the graded algebraic Bethe ansatz method.
Denote the nested double-row monodromy matrix as
\begin{eqnarray}
{\cal {T}}^{(1)}(\lambda ,\{ \mu _i\} )
\equiv T^{(1)}(\lambda , \{ \mu _i\} )
K^{(1)}(\lambda )
{T^{(1)}}^{-1}(-\lambda , 
\{ \mu _i\} )=\left( \begin{array}{cc}
{\cal {A}}^{(1)}(\lambda ) &{\cal {B}}^{(1)}(\lambda ) \\
{\cal {C}}^{(1)}(\lambda ) &{\cal {D}}^{(1)}(\lambda ) \end{array}
\right).
\end{eqnarray}
The nested double-row monodromy matrix satisfies
the reflection equation, 
\begin{eqnarray}
&&\hat {r}(\lambda -\mu )_{a_1a_2}^{b_1b_2}
{\cal {T}}^{(1)}(\lambda )_{b_1}^{c_1}
\hat {r}(\lambda +\mu )_{b_2c_1}^{c_2d_1}                           
{\cal {T}}^{(1)}(\mu )_{c_2}^{d_2}
(-)^{(\epsilon _{b_1}+\epsilon _{c_1})(1+\epsilon _{b_2})}
\nonumber \\
&=&{\cal {T}}^{(1)}(\mu )_{a_2}^{b_2}
\hat {r}(\lambda +\mu )_{a_1b_2}^{b_1c_2}
{\cal {T}}^{(1)}(\lambda )_{b_1}^{c_1}
\hat {r}(\lambda -\mu )_{c_2c_1}^{d_2d_1}                            
(-)^{(\epsilon _{b_1}+\epsilon _{c_1})(1+\epsilon _{c_2})}.
\end{eqnarray}
For convenience, we need the following transformation,
\begin{eqnarray}
{\cal {A}}^{(1)}(\lambda )=\tilde {\cal {A}}^{(1)}(\lambda )
-\frac {sin(\eta )e^{-2i\lambda }}{sin(2\lambda -\eta )}
{\cal {D}}^{(1)}(\lambda ).
\end{eqnarray}
With the help of this transformation, 
we have the following commutation relations:
\begin{eqnarray}
{\cal {D}}^{(1)}(\lambda )
{\cal {C}}^{(1)}(\mu )
&=&\frac {sin(\lambda -\mu +\eta )sin(\lambda +\mu )}
{sin(\lambda -\mu )sin(\lambda +\mu -\eta )}
{\cal {C}}^{(1)}(\mu )
{\cal {D}}^{(1)}(\lambda )
\nonumber \\
&-&\frac {sin(2\mu )sin(\eta )e^{i(\lambda -\mu )}}
{sin(\lambda -\mu )sin(2\mu -\eta )}
{\cal {C}}^{(1)}(\lambda )
{\cal {D}}^{(1)}(\mu )
+\frac {sin(\eta )e^{i(\lambda +\mu )}}{sin(\lambda +\mu -\eta )}
{\cal {C}}^{(1)}(\lambda )
\tilde {\cal {A}}^{(1)}(\mu ),
\end{eqnarray}
\begin{eqnarray}
\tilde {\cal {A}}^{(1)}(\lambda )
{\cal {C}}^{(1)}(\mu )
&=&\frac {sin(\lambda -\mu +\eta )sin(\lambda +\mu )}
{sin(\lambda -\mu )sin(\lambda +\mu -\eta )}
{\cal {C}}^{(1)}(\mu )
\tilde {\cal {A}}^{(1)}(\lambda )
\nonumber \\
&&-\frac {sin(\eta )sin(2\lambda )
e^{-i(\lambda -\mu )}}
{sin(\lambda -\mu )sin(2\lambda -\eta )}
{\cal {C}}^{(1)}(\lambda )
\tilde {\cal {A}}^{(1)}(\mu )
\nonumber \\
&&+\frac {sin(2\mu )sin(2\lambda )sin(\eta )e^{-i(\lambda +\mu )}}
{sin(\lambda +\mu -\eta )sin(2\lambda -\eta )sin(2\mu -\eta )}
{\cal {C}}^{(1)}(\lambda )
{\cal {D}}^{(1)}(\mu ),
\end{eqnarray}
\begin{eqnarray}
{\cal {C}}^{(1)}(\lambda )
{\cal {C}}^{(1)}(\mu )
= -\frac {sin(\lambda -\mu +\eta )}{sin(\lambda -\mu -\eta )} 
{\cal {C}}^{(1)}(\mu )
{\cal {C}}^{(1)}(\lambda  ).
\end{eqnarray}
In the same manner as the standard algebraic Bethe ansatz method,
acting the transfer matrix
\begin{eqnarray}
t^{(1)}(\lambda )
\equiv A^+(\lambda '+{\eta \over 2}){\cal {A}}^{(1)}(\lambda ')
-D^+(\lambda '+{\eta \over 2}){\cal {D}}^{(1)}(\lambda ')
=U_1^+(\lambda ')
\tilde {\cal {A}}^{(1)}(\lambda ')
-U_2^+(\lambda ')
{\cal {D}}^{(1)}(\tilde {\lambda '})
\end{eqnarray}
on the ansatz of the eigenvector
${\cal {C}}(\tilde {\mu }^{(1)}_1)
{\cal {C}}(\tilde {\mu }^{(1)}_2)\cdots 
{\cal {C}}(\tilde {\mu }^{(1)}_m)|0>^{(1)}$,
we find the eigenvalue of the nested transfer matrix as follows:
\begin{eqnarray}
\Lambda ^{(1)}(\lambda )
&=&U_1^+(\lambda ')
U_1(\lambda ')
\prod _{i=1}^n[sin(\lambda '+\mu '_i)
sin(\lambda '-\mu '_i)]
\prod _{l=1}^m \left\{
\frac {sin(\lambda '-\tilde {\mu }^{(1)}_l+\eta )
sin(\lambda '+\tilde {\mu }^{(1)}_l)}
{sin(\lambda '-\tilde {\mu }^{(1)}_l)
sin(\lambda '+\tilde {\mu }^{(1)}_l-\eta )}\right\} 
\nonumber\\ 
&&-U_2^+(\lambda ')
U_2(\lambda ')\prod _{i=1}^n
[sin(\lambda '+\mu '_i-\eta )
sin(\lambda '-\mu '_i-\eta )] 
\nonumber \\
&&\prod _{l=1}^m
\left\{ \frac {sin(\lambda '-\tilde {\mu }^{(1)}_l+\eta )
sin(\lambda '+\tilde {\mu }^{(1)}_l)}
{sin(\lambda '-\tilde {\mu }^{(1)}_l)
sin(\lambda '+\tilde {\mu }^{(1)}_l-\eta )}\right\} ,
\end{eqnarray}
where $\tilde {\mu }^{(1)}_1, \cdots ,\tilde {\mu }^{(1)}_m$
should satisfy the following Bethe ansatz equations,
\begin{eqnarray}
&&\frac {U_1^+(\tilde {\mu }_j^{(1)})
U_1(\tilde {\mu }_j^{(1)})}
{U_2^+(\tilde {\mu }_j^{(1)})
U_2(\tilde {\mu }_j^{(1)})}
\prod _{i=1}^n
\frac {sin(\tilde {\mu }_j^{(1)}+\mu '_i)
sin(\tilde {\mu }_j^{(1)}-\mu '_i)}
{sin(\tilde {\mu }_j^{(1)}+\mu '_i-\eta )
sin(\tilde {\mu }_j^{(1)}-\mu '_i-\eta )} 
=1,~~~j=1, \cdots , m.
\end{eqnarray}
The boundary parameters $U_1, U_1^+, U_2, U_2^+$ can be calculated
from the reflecting K-matrices (\ref{newk}, \ref{newdualk}). We
shall present the results in the next sub-section.

\subsection{Results for BFF grading}
In this sub-section, we shall summarize the results for the case of
BFF grading.
The eigenvalue of the transfer matrix $t(\lambda )$
with integrable boundary impurities is obtained as
\begin{eqnarray}
\Lambda (\lambda )
&=&-U_3^+(\lambda )U_3(\lambda )sin^{2N}(\lambda -\eta )
\prod _{i=1}^n
\frac {sin(\lambda +\mu _i)sin(\lambda -\mu _i+\eta )}
{sin(\lambda +\mu _i-\eta )sin(\lambda -\mu _i)}
\nonumber \\
&&+sin^{2N}(\lambda )
\prod _{i=1}^n
\frac {1}
{sin(\lambda -\mu _i)sin(\lambda +\mu _i-\eta )}
\Lambda ^{(1)}(\lambda ),
\end{eqnarray}
\begin{eqnarray}
\Lambda ^{(1)}(\lambda )
&=&U_1^+(\lambda )
U_1(\lambda )
\prod _{i=1}^n[sin(\lambda +\mu _i-\eta )
sin(\lambda -\mu _i)]
\prod _{l=1}^m \left\{
\frac {sin(\lambda -\mu ^{(1)}_l+\eta )
sin(\lambda +\mu ^{(1)}_l-\eta )}
{sin(\lambda -\mu ^{(1)}_l)
sin(\lambda +\mu ^{(1)}_l-2\eta )}\right\} 
\nonumber\\ 
&&-U_2^+(\lambda )
U_2(\lambda )\prod _{i=1}^n
[sin(\lambda +\mu _i-2\eta )
sin(\lambda -\mu _i-\eta )] 
\nonumber \\
&&\prod _{l=1}^m
\left\{ \frac {sin(\lambda -\mu ^{(1)}_l+\eta )
sin(\lambda +\mu ^{(1)}_l-\eta )}
{sin(\lambda -\mu ^{(1)}_l)
sin(\lambda +\mu ^{(1)}_l-2\eta )}\right\} ,
\end{eqnarray}
where $\mu _1^{(1)}, \cdots ,\mu _m^{(1)}$ and $\mu _1, \cdots , \mu _n$
should satisfy the Bethe ansatz equations:
\begin{eqnarray}
&&\frac {sin(2\mu _j)}
{sin(\mu _j-2\eta )}
\prod _{i=1,\not =j}^n \left\{ \frac 
{sin(\mu _j+\mu _i)sin(\mu _j-\mu _i+\eta )}
{sin(\mu _j+\mu _i-2\eta )
sin(\mu _j-\mu _i-\eta )}\right\}
\nonumber \\
&=&\frac
{sin^{2N}(\mu _j )}{sin^{2N}(\mu _j-\eta )}
\frac {U_2^+(\mu _j)U_2(\mu _j)}
{U_3^+(\mu _j)U_3(\mu _j)}
\prod _{l=1}^m
\left\{ \frac {sin(\mu _j-\mu ^{(1)}_l+\eta )
sin(\mu _j+\mu ^{(1)}_l-\eta )}
{sin(\mu _j-\mu ^{(1)}_l)
sin(\mu _j+\mu ^{(1)}_l-2\eta )}\right\} ,
\nonumber \\
&&~~~~j=1,\cdots ,n,
\end{eqnarray}
\begin{eqnarray}
&&\frac {U_1^+(\mu _j^{(1)})
U_1(\mu _j^{(1)})}
{U_2^+(\mu _j^{(1)})
U_2(\mu _j^{(1)})}
\prod _{i=1}^n
\frac {sin(\mu _j^{(1)}+\mu _i-\eta )
sin(\mu _j^{(1)}-\mu _i)}
{sin(\mu _j^{(1)}+\mu _i-2\eta )
sin(\mu _j^{(1)}-\mu _i-\eta )} 
=1,~~~j=1, \cdots , m.
\end{eqnarray}
Actually, these relations are rather general, we can take other
K-matrices. For the K-matrices considered in this paper, the
boundary parameters take the following form:
\begin{eqnarray}
U_1(\lambda )&=&
-\frac {sin(2\lambda )e^{-i(2\lambda +\eta )}}
{sin(2\lambda -2\eta )}
\frac {sin(\lambda +c-\eta )sin(\lambda -c)}
{sin(\lambda -c+\eta )sin(\lambda -c-\eta )},
\nonumber \\
U_2(\lambda )&=&
\frac {-e^{-2i\lambda }sin(2\lambda )sin(\lambda +c-2\eta )}
{sin(\lambda -c+\eta )sin(2\lambda -\eta )},
\nonumber \\
U_3(\lambda )&=&1,
\nonumber \\
U_1^+(\lambda )&=&
-\frac {e^{i(2\lambda +\eta )}sin(\lambda +\tilde {c}-{\eta \over 2})}
{sin(\lambda -\tilde {c}-{\eta \over 2})},
\nonumber \\
U_2^+(\lambda )&=&
\frac {-e^{2i\lambda }sin(2\lambda -\eta )
sin(\lambda -\tilde {c}-{3\eta \over 2})
sin(\lambda +\tilde {c}-{3\eta \over 2})}
{sin(2\lambda -2\eta )sin^2(\lambda -\tilde {c}-{\eta \over 2})}
\nonumber \\
U_3^+(\lambda )&=&\frac {sin(\lambda -\tilde {c}+{\eta \over 2})
sin(\lambda -\tilde {c}-{3\eta \over 2})}
{sin^2(\lambda -\tilde {c}-{\eta \over 2})}.
\end{eqnarray}
The energy of the Hamiltonian is given by
\begin{eqnarray}
E=-Ncos(\eta )-\sum _{j=1}^n\frac {sin^2(\eta )}
{sin(\mu _j)sin(\mu _j-\eta )}
-\frac {sin^3(\eta )cos(\tilde {c}+{\eta \over 2})}
{sin(\tilde {c}-{\eta \over 2})
sin(\tilde {c}+{\eta \over 2})sin(\tilde {c}+{3\eta \over 2})}.
\end{eqnarray}

\section{Results of the FBF grading}
We have dealt with the gradings FFB and BFF.
We shall study
the last possible grading, FBF, $\epsilon _1=\epsilon _3=1,
\epsilon _2=0$. In this case, we choose the following
representation,
\begin{eqnarray}
&&S_k=e^k_{13}, ~~~S_k^{\dagger }=e^k_{31},
~~~S_k^z={1\over 2}(e^k_{11}-e^k_{33}),
\nonumber \\
&&Q_{k,1}=(1-n_{k,-1})c_{k,1}=e^k_{21},
~~~~Q_{k,1}^{\dagger }=(1-n_{k,-1})c_{k,1}^{\dagger }=e^k_{12},
~~~~Q_{k,-1}=(1-n_{k,1})c_{k,-1}=e^k_{23},
\nonumber \\
&&Q_{k,-1}^{\dagger }=(1-n_{k,1})c_{k,-1}^{\dagger }=e^k_{32},
~~~~T_k=1-{1\over 2}n_k={1\over 2}(e^k_{33}+e^k_{11})+e^k_{22}.
\end{eqnarray}
The calculation can be preformed in the same way as the BFF
grading. Here we present some main results.
We still suppose that K-matrix operators satisfying the reflection equation
and the dual reflection equation take the form
(\ref{K1},\ref{dualK}). We have
the following solution to the reflection equation
\begin{eqnarray}
A(\lambda )&=&{1\over 2}
g(\lambda )\left[ sin(\lambda +c-\eta )
sin(\lambda -c-\eta )+sin(\lambda +c)sin(\lambda -c)
\right. \nonumber \\
&&\left.
+sin^2(\eta )e^{-2i\lambda }-sin({\eta })sin(2\lambda )e^{-i\eta }\sigma ^z
\right] ,
\nonumber \\
B(\lambda )&=&g(\lambda )sin(\eta )sin(2\lambda )
\sigma ^-,
\nonumber \\
C(\lambda )&=&g(\lambda )sin(\eta )sin(2\lambda )
\sigma ^+,
\nonumber \\
D(\lambda )&=&{1\over 2}g(\lambda )
\left[ sin(\lambda +c+\eta )
sin(\lambda -c+\eta )+sin(\lambda +c)sin(\lambda -c)
\right. \nonumber \\
&&\left. +sin^2(\eta )e^{2i\lambda }
-sin(\eta )sin(2\lambda )e^{-i\eta }\sigma ^z\right],
\end{eqnarray}
where
\begin{eqnarray}
g(\lambda )=\frac {-e^{-2i\lambda }}
{sin(\lambda -c+\eta )sin(\lambda -c-\eta )}.
\end{eqnarray}
For the case of the dual reflection equation, we have
\begin{eqnarray}
A^+(\lambda )&=&{1\over 2}
g^+(\lambda )\left[ sin(\lambda +\tilde {c}-{3\eta \over 2})
sin(\lambda -\tilde {c}-{3\eta \over 2})
+sin(\lambda +\tilde {c}-{\eta \over 2})
sin(\lambda -\tilde {c}-{\eta \over 2})
\right. \nonumber \\
&&\left.
-sin^2(\eta )e^{i(2\lambda -\eta )}
-sin({\eta })sin(2\lambda -\eta )e^{i\eta }\tilde {\sigma }^z
\right] ,
\nonumber \\
B^+(\lambda )&=&g^+(\lambda )e^{i\eta }sin(\eta )sin(2\lambda -\eta )
\tilde {\sigma }^-,
\nonumber \\
C^+(\lambda )&=&g^+(\lambda )e^{i\eta }sin(\eta )sin(2\lambda -\eta )
\tilde {\sigma }^+,
\nonumber \\
D^+(\lambda )&=&{1\over 2}g^+(\lambda )
\left[ sin(\lambda +\tilde {c}+{\eta \over 2})
sin(\lambda -\tilde {c}+{\eta \over 2})
+sin(\lambda +\tilde {c}-{\eta \over 2})
sin(\lambda -\tilde {c}-{\eta \over 2})
\right. \nonumber \\
&&\left. -sin^2(\eta )e^{-i(2\lambda -\eta )}
-sin(\eta )sin(2\lambda -\eta )e^{i\eta }
\tilde {\sigma }^z\right],
\end{eqnarray}
where
\begin{eqnarray}
g^+(\lambda )=\frac {-e^{i(2\lambda -\eta )}}
{sin^2(\lambda -\tilde {c}-{\eta \over 2})}.
\end{eqnarray}
The boundary impurity term in the Hamiltonian defined
by the K-matrix is written as
\begin{eqnarray}
H_1&=&\frac {2sin(\eta )}{sin(\eta +c)sin(\eta -c)}
(1-n_{1,-1})[a_L^{\dagger }c_{1,-1}
+c_{1,-1}^{\dagger }a_L-{\frac {e^{-i\eta }}{2}}(1-2n_L)]
\nonumber \\
&&+(1-n_{1,-1})
\left( \frac {2sin(2c)}{sin(\eta +c)sin(c-\eta )}-4i\right)
+\frac {sin(\eta )e^{i\eta }}{sin(\eta +c)sin(\eta -c)}
(3T_1-S_1^z-2).
\end{eqnarray}
The eigenvalue of the transfer matrix with
boundary impurities is obtained as
\begin{eqnarray}
\Lambda (\lambda )
&=&-U_3^+(\lambda )U_3(\lambda )sin^{2N}(\lambda -\eta )
\prod _{i=1}^n
\frac {sin(\lambda +\mu _i)sin(\lambda -\mu _i+\eta )}
{sin(\lambda +\mu _i-\eta )sin(\lambda -\mu _i)}
\nonumber \\
&&+sin^{2N}(\lambda )
\prod _{i=1}^n
\frac {1}
{sin(\lambda -\mu _i)sin(\lambda +\mu _i-\eta )}
\Lambda ^{(1)}(\lambda ),
\nonumber \\
\Lambda ^{(1)}(\lambda )
&=&-U_1^+(\lambda )
U_1(\lambda )
\prod _{i=1}^n[sin(\lambda +\mu _i-\eta )
sin(\lambda -\mu _i)]
\prod _{l=1}^m \left\{
\frac {sin(\lambda -\mu ^{(1)}_l-\eta )
sin(\lambda +\mu ^{(1)}_l-\eta )}
{sin(\lambda -\mu ^{(1)}_l)
sin(\lambda +\mu ^{(1)}_l)}\right\} 
\nonumber\\ 
&&+U_2^+(\lambda )
U_2(\lambda )\prod _{i=1}^n
[sin(\lambda +\mu _i)
sin(\lambda -\mu _i+\eta )] 
\nonumber \\
&&\prod _{l=1}^m
\left\{ \frac {sin(\lambda -\mu ^{(1)}_l-\eta )
sin(\lambda +\mu ^{(1)}_l-\eta )}
{sin(\lambda -\mu ^{(1)}_l)
sin(\lambda +\mu ^{(1)}_l)}\right\} ,
\end{eqnarray}
where $\mu ^{(1)}_1, \cdots ,\mu ^{(1)}_m$
should satisfy the following Bethe ansatz equations,
\begin{eqnarray}
&&\frac {U_1^+(\mu _j^{(1)})
U_1(\mu _j^{(1)})}
{U_2^+(\mu _j^{(1)})
U_2(\mu _j^{(1)})}
\prod _{i=1}^n
\frac {sin(\mu _j^{(1)}+\mu _i-\eta )
sin(\mu _j^{(1)}-\mu _i)}
{sin(\mu _j^{(1)}+\mu _i)
sin(\mu _j^{(1)}-\mu _i+\eta )} 
=1,~~~j=1, \cdots , m,
\end{eqnarray}
and $\mu _1, \cdots, \mu _n$ should satisfy 
\begin{eqnarray}
1=\frac
{sin^{2N}(\mu _j )}{sin^{2N}(\mu _j-\eta )}
\frac {U_2^+(\mu _j)U_2(\mu _j)}
{U_3^+(\mu _j)U_3(\mu _j)}
\prod _{l=1}^m
\left\{ \frac {sin(\mu _j-\mu ^{(1)}_l-\eta )
sin(\mu _j+\mu ^{(1)}_l-\eta )}
{sin(\mu _j-\mu ^{(1)}_l)
sin(\mu _j+\mu ^{(1)}_l)}\right\} ,
~~~~j=1,\cdots ,n.
\end{eqnarray}
The boundary parameters are as follows:
\begin{eqnarray}
U_1(\lambda )&=&
\frac {-e^{-i(2\lambda -\eta )}sin(\lambda -c)sin(\lambda +c-\eta )}
{sin(\lambda -c+\eta )sin(\lambda -c-\eta )},
\nonumber \\
U_2(\lambda )&=&
\frac {-e^{-2i\lambda }sin(2\lambda )sin(\lambda +c)}
{sin(\lambda -c-\eta )sin(2\lambda -\eta )},
\nonumber \\
U_3(\lambda )&=&1,
\nonumber \\
U_1^+(\lambda )
&=&\frac {-e^{i(2\lambda -\eta )}sin(\lambda +\tilde {c}-{\eta \over 2})}
{sin(\lambda -\tilde {c}-{\eta \over 2})},
\nonumber \\
U_2^+(\lambda )&=&
\frac {-e^{2i\lambda }sin(2\lambda -\eta )
sin(\lambda -\tilde {c}+{\eta \over 2})
sin(\lambda +\tilde {c}+{\eta \over 2})}
{sin(2\lambda )
sin^2(\lambda -\tilde {c}-{\eta \over 2})},
\nonumber \\
U_3^+(\lambda )&=&
\frac {sin(\lambda -\tilde {c}+{\eta \over 2})
sin(\lambda -\tilde {c}-{3\eta \over 2})}
{sin^2(\lambda -\tilde {c}-{\eta \over 2})}.
\end{eqnarray}
The energy of the Hamiltonian is the same as that of the BFF grading,
\begin{eqnarray}
E=-Ncos(\eta )-\sum _{j=1}^n\frac {sin^2(\eta )}
{sin(\mu _j)sin(\mu _j-\eta )}
-\frac {sin^3(\eta )cos(\tilde {c}+{\eta \over 2})}
{sin(\tilde {c}-{\eta \over 2})
sin(\tilde {c}+{\eta \over 2})sin(\tilde {c}+{3\eta \over 2})}.
\end{eqnarray}

\section{Summary}
In this paper, we have studied the integrable boundary impurity
problem for the generalized (q-deformed) supersymmetric
$t-J$ model in all possible three gradings, FFB, BFF and
FBF. We have presented reflecting K-matrix opertors which are
solutions to the reflection equation in different gradings.
Using the nested algebraic Bethe ansatz method, we have obtained
the eigenvalues for the transfer matrix and
the Hamiltonain with integrable boundary impurities.

In all three possible gradings, we suppose that
reflecting K-matrices take a similar form (\ref{K},\ref{K1}).
We remark that K-matrix remains the same form while R-matrix
changes for different gradings. Thus the boundary terms in
the Hamiltonian for different gradings are completely different.
Each grading may correspond to an
integrable boundary impurity. It is interesting to analyze
the Bethe ansatz equations and compare the
ground state properties, low-lying excitations and
the thermodynamic limit for different gradings.
These remain as our future problems.

\vskip 1truecm
\noindent {\bf Acknowlegements:} One of the authors, H.F. acknowleges the
support of JSPS. The authors also thank H.Frahm, R.H.Yue
and H.Q.Zhou for useful discussions. This work was partly supported by
NSFC, Clibming project and NWU teachers fund.

\end{document}